\documentclass[twocolumn,preprintnumbers,amsmath,amssymb,floatfix]{revtex4}
\usepackage{graphicx}
\usepackage{dcolumn}
\usepackage{bm}
\usepackage{epsf}
\usepackage{amsmath}
\usepackage{amsfonts}

\renewcommand{\apj}{ApJ}
\newcommand{\apjl}{ApJ}

\begin{document}
\title{An exact integral relation between the $^{56}$Ni mass and the bolometric light curve of a type Ia supernova}
\author{Boaz Katz$^{1*}$, Doron Kushnir$^1$, Subo Dong$^1$}
\affiliation{$^1$Institute for Advanced Study, Princeton, NJ 08540, USA}
\begin{abstract}
\end{abstract}
\maketitle

The mass of $^{56}$Ni that is formed at the onset of a type Ia supernova is an essential parameter in its modeling. Significant effort has been invested in deriving approximate relations between the light curves and this mass (e.g. \cite{Colgate69,Arnett79,Colage80,Arnett82,Pinto00}). An exact relation between the $^{56}$Ni mass and the bolometric light curve can be derived as follows, using the following excellent approximations: 1. the emission is powered solely by $^{56}$Ni $\rightarrow~^{56}$Co $\rightarrow~^{56}$Fe;  2. each mass element propagates at a non-relativistic velocity which is  constant in time (free coasting); and 3. the internal energy is dominated by radiation. Under these approximations, the energy $E(t)$ carried by radiation in the ejecta satisfies: 
\begin{equation}
dE/dt=-E(t)/t-L_{\rm bol}(t)+Q(t),
\end{equation} 
where $Q(t)=Q_{\rm Ni}M_{\rm Ni}(t)/\tau_{\rm Ni}+Q_{\rm Co}M_{\rm Co}(t)/\tau_{\rm Co}$ is the deposition of energy by the decay which is precisely known. Multiplying this relation by $t$ and integrating over time we find: $E(t)\cdot t=\int_0^t Q(t')~t'~dt' -\int_0^t L_{\rm bol}(t')~t'~dt'$. At late times, $t\gg t_{\rm peak}$, the energy inside the ejecta decreases rapidly due to its escape, and thus we have 
\begin{equation}\label{eq:result}
\int_0^t Q(t')~t'~dt'=\int_0^t L_{\rm bol}(t')~t'~dt'~~~~~~~~\rm{at}~~t\gg t_{\rm peak}.
\end{equation}
The right hand side of eq. \eqref{eq:result} is an observable while the left hand side is a known function that is proportional to the mass of $^{56}$Ni formed at the explosion. 
We emphasize that this relation is correct regardless of the opacities, density distribution or $^{56}$Ni deposition distribution in the ejecta and is very different from ``Arnett's rule'', $L_{\rm peak}\sim Q(t_{\rm peak})$ \cite{Arnett79,Arnett82}. By comparing $\int_0^t Q(t')~t'~dt'$ with $\int_0^t L_{\rm bol}(t')~t'~dt'$ at $t \sim 40\rm ~day$ after the explosion, the mass of $^{56}$Ni can be found directly from UV, optical and infrared observations with modest corrections due to the unobserved gamma-rays and due to the small residual energy in the ejecta, $E(t)\cdot t>0$. 

B.K. was supported by NASA through Einstein Postdoctoral Fellowship awarded by the Chandra X-ray Center, which is operated by the Smithsonian Astrophysical Observatory for NASA under contract NAS8-03060. D.K. was supported by NSF grant AST-0807444. S.D. was supported through a Ralph E. and Doris M. Hansmann Membership at the IAS and NSF grant AST-0807444.


\begin{thebibliography}{99}


\bibitem[Colgate 
\& McKee(1969)]{Colgate69} Colgate, S.~A., \& McKee, C.\ 1969, \apj, 157, 623 

\bibitem[Arnett(1979)]{Arnett79} Arnett, W.~D.\ 1979, \apjl, 
230, L37

\bibitem[Colgate et al.(1980)]{Colage80} Colgate, S.~A., 
Petschek, A.~G., \& Kriese, J.~T.\ 1980, \apjl, 237, L81 


\bibitem[Arnett(1982)]{Arnett82} Arnett, W.~D.\ 1982, \apj, 253, 
785 

\bibitem[Pinto 
\& Eastman(2000)]{Pinto00} Pinto, P.~A., \& Eastman, R.~G.\ 2000, \apj, 530, 744 

*Bahcal Fellow, Einstein Fellow
\end{thebibliography}
\end{document}